%% file: Two-Qubit-Additivity.tex
\documentclass[aps,prl,reprint,longbibliography,floatfix,superscriptaddress]{revtex4-2}
\usepackage{amsmath,amssymb,amsthm,mathtools}
\usepackage{bm,graphicx}
\usepackage{times}
\usepackage{xcolor,etoolbox}
\usepackage[colorlinks=true,citecolor=blue,linkcolor=blue,urlcolor=blue]{hyperref}
\usepackage{bookmark}
\usepackage{zref-abspage}

\theoremstyle{definition}
\newtheorem{theorem}{Theorem}
\newtheorem{lemma}{Lemma}

\AtBeginEnvironment{thebibliography}{\interlinepenalty=10000\relax}

\begin{document}
\newcommand{\manuscripttitle}{Entanglement Cost of an Arbitrary State of Two Qubits}
\hypersetup{pdftitle={\manuscripttitle}}
\title{\manuscripttitle}
\author{Dong-Dong Dong}
\affiliation{School of Materials Science and Engineering, and School of Physics, Anhui University, Hefei 230601, People's Republic of China}
\author{Xue-Ke Song}
\affiliation{School of Materials Science and Engineering, and School of Physics, Anhui University, Hefei 230601, People's Republic of China}
\author{Liu Ye}
\affiliation{School of Materials Science and Engineering, and School of Physics, Anhui University, Hefei 230601, People's Republic of China}

\author{Gerardo Adesso}
\email{gerardo.adesso@nottingham.ac.uk}
\affiliation{School of Mathematical Sciences and Centre for the Mathematics and Theoretical Physics of Quantum Non-Equilibrium Systems, University of Nottingham, University Park, Nottingham NG7 2RD, United Kingdom}
\affiliation{Algorithmiq S.r.L., Via della Chiusa 15, 20123 Milano, Italy}
\author{Dong Wang}
\email{dwang@ahu.edu.cn}
\affiliation{School of Materials Science and Engineering, and School of Physics, Anhui University, Hefei 230601, People's Republic of China}
\date{November 18, 2026}

\begin{abstract}
We prove that the entanglement cost of every two-qubit state equals its entanglement of formation (EOF), as given by Wootters' formula. We establish this by proving strong superadditivity of EOF for a qubit pair across a finite-dimensional cut. This yields additivity with any finite-dimensional bipartite auxiliary state and a sharp lower bound on multiqubit cost from disjoint pairs. We find the exact parallel simulation cost of all qubit-to-qubit channels. We further derive a single-use formula for the energy-constrained classical capacity of all channels whose output and environment are qubits, including those with three- and four-dimensional inputs.
\end{abstract}
\maketitle
\raggedbottom
\begingroup
\emph{Introduction.}---Consider two distant laboratories that use local operations and classical communication (LOCC) to prepare many independent copies of a state $\rho$. Their entanglement cost $E_C(\rho)$ is the infimum of the asymptotic number of Bell pairs consumed per copy as the preparation error tends to zero. A preparation strategy starts from an ensemble of pure states realizing $\rho$. Preparing its members at their pure-state entanglement rates and minimizing the average gives the entanglement of formation (EOF)~\cite{Bennett1996,HaydenHorodeckiTerhal2001}. Preparing blocks of $k$ copies allows the ensemble states to be entangled across copies. Whether that freedom lowers the rate is the question of interest here.

For a bipartite density operator $\rho$ on $A\mid B$ we write its EOF as
\begin{equation}
 E_F(\rho)=\inf_{\{q_l,\psi_l\}}\sum_l q_l S\bigl(\operatorname{Tr}_B|\psi_l\rangle\langle\psi_l|\bigr),
 \label{eq:roof}
\end{equation}
where the infimum runs over normalized pure-state ensembles of $\rho$ and $S$ is the von Neumann entropy in bits. Hayden \textit{et al.} identified the operational preparation cost with the regularized EOF~\cite{HaydenHorodeckiTerhal2001},
\begin{equation}
 E_C(\rho)=\lim_{k\to\infty}\frac{E_F(\rho^{\otimes k})}{k}.
 \label{eof:entanglement-cost}
\end{equation}
For two qubits, Wootters evaluated the single-copy convex roof exactly~\cite{Wootters1998}. Using this value for the rate in Eq.~\eqref{eof:entanglement-cost} requires control of ensembles of arbitrarily many copies.

Whether EOF is additive for every pair of bipartite states was listed as Open Quantum Problem~7 in the collection edited by Kr{\"u}ger and Werner~\cite{KrugerWerner2005}. Shor established the equivalence of the unrestricted EOF and channel additivity conjectures~\cite{Shor2004}, and Hastings' counterexample rules out universal additivity~\cite{Hastings2009}.
Pomeransky showed that additivity of EOF for all bipartite states would itself imply strong superadditivity~\cite{Pomeransky2003}. For four qubits, Cornelio and de Oliveira found numerical violations of strong superadditivity for R\'enyi entanglement at orders above one, but observed none at order one, where the measure reduces to EOF~\cite{CornelioDeOliveira2010}.

Vidal \textit{et al.} established EOF additivity with an arbitrary auxiliary state for certain support subspaces, including those relevant to mixtures of two Bell states~\cite{VidalDurCirac2002}. Fang obtained exact costs for two-qubit states whose normalized negativity equals their concurrence, a class containing all Bell-diagonal states~\cite{Fang2026}. In continuous variables, Adesso established EOF additivity and $E_C=E_F$ for two-mode Gaussian states~\cite{Adesso2026GaussianEOF}, and Gao and Adesso extended these results to finite-energy multimode Gaussian states and proved strong superadditivity~\cite{GaoAdesso2026Multimode}. They also obtained the simulation cost and energy-constrained classical capacity of multimode Gaussian channels from covariance matrices. For a selected qubit pair, Tang \textit{et al.} proved strong superadditivity whenever the kernel of its reduced state contains a nonzero product vector~\cite{TangZhuBaiWang2026}. Their result allows an arbitrary finite-dimensional bipartite remainder and covers every two-qubit marginal of rank at most two.

In this Letter, we prove strong superadditivity for a two-qubit marginal of arbitrary rank and support. Applied to product states, the theorem gives additivity with any finite-dimensional bipartite auxiliary state and determines the exact entanglement cost of every two-qubit state. Iteration over disjoint pairs gives a sharp lower bound on the preparation cost of a multiqubit state. The theorem also determines the cost of simulating qubit channels in parallel and the energy-constrained classical capacity of channels with a qubit output and a qubit environment.

\emph{Strong superadditivity.}---Alice holds $A_1A'$ and Bob holds $B_1B'$, where $A_1$ and $B_1$ are the selected qubits.
\begin{theorem}[Strong superadditivity]\label{eof:strong-superadditivity}
For every state $\rho$ on $A_1A'B_1B'$ with $A_1,B_1$ qubits and $A',B'$ finite-dimensional,
\begin{equation}
 E_F(\rho_{A_1A'\mid B_1B'})\ge E_F(\rho_{A_1B_1})+E_F(\rho_{A'B'}).
 \label{eof:twoqubit-strong-superadditivity}
\end{equation}
\end{theorem}
{\emph{Sketch of proof.}---Suppose first that $\rho$ is pure and let $\omega=\rho_{A_1B_1}$ be the selected pair's marginal. Local rank-one measurements on $A_1$ and $B_1$ leave a pure state on $A'B'$ for each outcome of nonzero probability. Write $p_{xy}$ for the probability of outcome $(x,y)$, with $x,y\in\{0,1\}$, and $\rho_{A'}^{xy}$ for the marginal on $A'$ in each nonzero branch. Averaging the conditional states with weights $p_{xy}$ gives $\rho_{A'B'}$, so the definition of EOF yields
\[
 \sum_{x,y}p_{xy}S(\rho_{A'}^{xy})\ge E_F(\rho_{A'B'}),
\]
with zero-probability terms omitted. Define the mean entanglement loss by
\[
 \Delta=S(\rho_{A_1A'})-\sum_{x,y}p_{xy}S(\rho_{A'}^{xy}).
\]
To prove Eq.~\eqref{eof:twoqubit-strong-superadditivity}, it therefore suffices to choose measurements with $\Delta\ge E_F(\omega)$.

For measurements with unbiased outcomes on each qubit, we first show that $\Delta\ge W(|t|)$, where $t$ is the mean product of the two measurement eigenvalues and $W(c)$ is the two-qubit EOF at concurrence $c$. We then choose the directions so that $W(|t|)\ge E_F(\omega)$.

Grouping the four outcome blocks by parity and differentiating the Schatten Hanner inequality~\cite{Heinavaara2025} at exponent two gives the four-block entropy estimate in Eq.~\eqref{app:match-general-loss}. For unbiased measurements, the weighted entropy within the parity groups is one bit and cancels the subtracted bit, yielding $\Delta\ge W(|t|)$ in Eq.~\eqref{app:match-unbiased-loss}.

Using only the pair marginal $\omega$, Lemma~\ref{app:match-transverse} supplies unbiased measurement directions with $|t|\ge\mathcal C(\omega)$. The monotonicity of $W$~\cite{Wootters1998} then gives
\[
 \Delta\ge W(|t|)\ge E_F(\omega).
\]
The bound is attained at every concurrence by a pure qubit pair in product with a pure remainder. Adding the lower bounds on $\Delta$ and on the average branch entanglement gives Eq.~\eqref{eof:twoqubit-strong-superadditivity} for pure states.

For mixed $\rho$, apply Eq.~\eqref{eof:twoqubit-strong-superadditivity} to each state in any pure-state decomposition of $\rho$ and average the resulting inequalities. Convexity on both marginals and the infimum over global ensembles give Eq.~\eqref{eof:twoqubit-strong-superadditivity}. Thus tracing out even a correlated pair reduces the global EOF by at least the pair's EOF. The full proof is given in Sec.~\ref{app:eof-strong-superadditivity} of the Supplemental Material~\cite{PRLSupplementalMaterial}.\qed}

Apply Theorem~\ref{eof:strong-superadditivity} to a product $\rho\otimes\sigma$, where $\rho$ is any two-qubit state and $\sigma$ any finite-dimensional bipartite state. Its marginals are $\rho$ and $\sigma$. By entropy additivity, the product of optimal pure-state ensembles for $\rho$ and $\sigma$ has average entanglement $E_F(\rho)+E_F(\sigma)$, so Eq.~\eqref{eq:roof} gives
\[
 E_F(\rho\otimes\sigma)\le E_F(\rho)+E_F(\sigma).
\]
Theorem~\ref{eof:strong-superadditivity} gives the opposite inequality for these marginals, so
\begin{equation}
 E_F(\rho\otimes\sigma)=E_F(\rho)+E_F(\sigma).
 \label{eq:full-additivity}
\end{equation}
Choosing $\sigma=\rho^{\otimes(k-1)}$ and iterating gives $E_F(\rho^{\otimes k})=kE_F(\rho)$ for every positive integer $k$. Equation~\eqref{eof:entanglement-cost} now gives a closed expression for the operational rate,
\begin{equation}
 \begin{gathered} E_C(\rho)=E_F(\rho)=W(\mathcal C(\rho)),\\
 W(c)=h_2\!\left(\frac{1+\sqrt{1-c^2}}{2}\right),
 \end{gathered}\label{eq:cost}
\end{equation}
where $h_2$ is the binary entropy. The concurrence $\mathcal C(\rho)$ is $\max(0,\lambda_1-\lambda_2-\lambda_3-\lambda_4)$, where $\lambda_1\ge\lambda_2\ge\lambda_3\ge\lambda_4$ are the nonnegative square roots of the eigenvalues of $\rho\widetilde\rho$ and $\widetilde\rho=(\sigma_y\otimes\sigma_y)\rho^*(\sigma_y\otimes\sigma_y)$ in the computational basis~\cite{Wootters1998}. The asymptotic entanglement cost of any two-qubit state $\rho$ is therefore determined by the four eigenvalues of $\rho\widetilde\rho$, the product of $\rho$ and its spin-flipped state.

To illustrate the scope of Eq.~\eqref{eq:cost}, consider a state outside the conditions of both Tang \textit{et al.}~\cite{TangZhuBaiWang2026} and Fang~\cite{Fang2026}. With $|\Phi^+\rangle=(|00\rangle+|11\rangle)/\sqrt{2}$, take
\[
 \rho_\star=\tfrac25\bigl(|\Phi^+\rangle\langle\Phi^+|+|01\rangle\langle01|\bigr)+I_4/20.
\]
Its smallest eigenvalue is $1/20$, so it has full rank and its kernel contains no nonzero product vector. Its concurrence is $\mathcal C(\rho_\star)=1/10$, whereas its normalized negativity is $\|\rho_\star^{T_B}\|_1-1=(4\sqrt{2}-5)/10<1/10$, ruling out Fang's equality condition. Equation~\eqref{eq:cost} gives the exact entanglement cost $E_C(\rho_\star)=W(1/10)$.

\emph{Entanglement-cost bound from disjoint pairs.}---Suppose Alice and Bob each hold several qubits. An experimental estimate of the full density matrix may be costly, whereas the reduction of each cross pair requires only local one- and two-qubit Pauli averages. For any multiqubit state $\rho$ across $A\mid B$, these pair marginals give the bound
\begin{equation}
 E_F(\rho_{A\mid B})\ge E_C(\rho_{A\mid B})
 \ge\max_M\sum_{(i,j)\in M}E_F(\rho_{A_iB_j}),
 \label{eof:matching-bound}
\end{equation}
where $M$ ranges over sets of cross pairs sharing no qubit.

Iterating Eq.~\eqref{eof:twoqubit-strong-superadditivity} over the disjoint pairs in $M$ gives the EOF bound. At each step, tracing out one pair leaves the marginals of all remaining pairs unchanged. In $\rho^{\otimes k}$, choose the same matching in each copy. Applying the EOF bound to all $k|M|$ disjoint pairs and dividing by $k$ gives
\[
 \frac{E_F(\rho^{\otimes k})}{k}
 \ge \sum_{(i,j)\in M} E_F(\rho_{A_iB_j}).
\]
Taking $k\to\infty$ in Eq.~\eqref{eof:entanglement-cost} gives the cost bound, with arbitrary correlations among the pairs within each copy. Evaluating the maximum in Eq.~\eqref{eof:matching-bound} is a standard maximum-weight bipartite matching problem.

The bound is sharp even if the full density matrices of the selected pairs are given. For any prescribed two-qubit states $\rho_{A_iB_j}$ on the disjoint pairs in $M$, let $\rho_{\mathrm{prod}}$ be their tensor product with a separable state on the unpaired qubits. This state has the prescribed pair marginals. Iterating Eq.~\eqref{eq:full-additivity} gives $E_F(\rho_{\mathrm{prod}})$ equal to the sum of the prescribed EOFs. The maximum in Eq.~\eqref{eof:matching-bound} is at least this sum because $M$ is an admissible matching. Hence $E_C(\rho_{\mathrm{prod}})=E_F(\rho_{\mathrm{prod}})$, and the matching bound attains the same value. If a mixture of such products induces an EOF-optimal pure-state ensemble for each selected pair, the average EOF of its components equals the sum of the pair EOFs. By convexity, this sum upper-bounds the global EOF even when the preparation labels are classically correlated, so both inequalities in Eq.~\eqref{eof:matching-bound} hold with equality.

Squashed entanglement $E_{\mathrm{sq}}$ gives a corresponding matching bound by strong superadditivity and $E_{\mathrm{sq}}\le E_C$~\cite{ChristandlWinter2004}. For the same pair marginals, Eq.~\eqref{eof:matching-bound} is at least as tight because $E_{\mathrm{sq}}\le E_F$ on each pair, with the pair EOFs given directly by Wootters' formula~\cite{Wootters1998}. The four-qubit state $(|0000\rangle+|1111\rangle)/\sqrt{2}$ illustrates the information lost by looking only at pairs. Across a two-versus-two cut its preparation cost is one ebit, yet every cross-pair marginal is a separable mixture of $|00\rangle$ and $|11\rangle$. Equation~\eqref{eof:matching-bound} then returns zero.

\emph{Channel simulation cost and classical capacity.}---Consider first a channel $\mathcal N$ with qubit input and output. Its parallel LOCC simulation cost $E_C^{\rm par}(\mathcal N)$ is the minimum asymptotic number of Bell pairs per use needed to simulate $\mathcal N^{\otimes n}$ on arbitrary inputs with vanishing diamond-norm error~\cite{Berta2013Channels}. Let $J_{\mathcal N}=(\mathrm{id}\otimes\mathcal N)(|\Phi^+\rangle\langle\Phi^+|)$ be its normalized Choi state. The known bounds $E_C(J_{\mathcal N})\le E_C^{\rm par}(\mathcal N)\le E_F(J_{\mathcal N})$~\cite{TangZhuBaiWang2026} coincide by Eq.~\eqref{eq:cost}, giving
\begin{equation}
 E_C^{\rm par}(\mathcal N)=W\bigl(\mathcal C(J_{\mathcal N})\bigr).
 \label{eq:channel-simulation}
\end{equation}
Tang \textit{et al.} obtained this equality for qubit channels admitting a pure output~\cite{TangZhuBaiWang2026}, and Fang evaluated the cost of qubit depolarizing channels~\cite{Fang2026}. Equation~\eqref{eq:channel-simulation} holds for every qubit-to-qubit channel. In particular, it gives the parallel simulation cost of finite-temperature generalized amplitude-damping channels~\cite{KhatriSharmaWilde2020}, whose unconstrained classical capacity was determined by Fang~\cite{Fang2026GADC}.

For classical communication, the relevant qubits are the output $B$ and the environment $E$. Suppose $\mathcal N(\rho)=\operatorname{Tr}_E(V\rho V^\dagger)$ for an isometry $V:X\to B\otimes E$ with both $B$ and $E$ qubits. The input $X$ can therefore have dimension up to four. For an input Hamiltonian $H$ and a feasible mean energy budget $e$ per use, the unassisted classical capacity in bits per use is
\begin{equation}
 C_{\rm cl}(\mathcal N;H,e)=
 \max_{\rho:\,\operatorname{Tr}(H\rho)\le e}
 \bigl[S(\mathcal N(\rho))-E_F(V\rho V^\dagger)\bigr].
 \label{eq:channel-capacity}
\end{equation}
Here $\rho$ ranges over single-use input states, and Wootters' formula gives the EOF across $B\mid E$. At fixed $\rho$, the expression in brackets equals $\chi_{\mathcal N}(\rho)$, the maximum Holevo information over input ensembles with mean $\rho$~\cite{MSW2004}.

To see why joint encoding cannot raise the rate, let $\tau$ be the average input for $n$ uses, with marginals $\tau_i$ and $\bar\tau=n^{-1}\sum_i\tau_i$. Across $B^n\mid E^n$, iterating Theorem~\ref{eof:strong-superadditivity} over the pairs $B_iE_i$ gives
\[
 E_F\!\left(V^{\otimes n}\tau(V^\dagger)^{\otimes n}\right)
 \ge \sum_{i=1}^n E_F(V\tau_i V^\dagger).
\]
Combining this bound with subadditivity of the output entropy and concavity of $\chi_{\mathcal N}$ in the mean input gives
\[
 \chi_{\mathcal N^{\otimes n}}(\tau)
 \le \sum_{i=1}^n\chi_{\mathcal N}(\tau_i)
 \le n\,\chi_{\mathcal N}(\bar\tau).
\]
The energy constraint gives $\operatorname{Tr}(H\bar\tau)\le e$, so Eq.~\eqref{eq:channel-capacity} upper-bounds the rate per use. Product signals attain this rate by the coding theorem for constrained inputs~\cite{Holevo1998Coding}.

For qubit $B,E$, the support criterion of Tang \textit{et al.} assumes that the orthogonal complement of the image of $V$ contains a product vector~\cite{TangZhuBaiWang2026}. Our proof applies to every isometry into $B\otimes E$. Wu's fixed-input-marginal bound yields a single-use capacity formula under additive input-energy constraints for all qubit-to-qubit channels and for certain channels with higher-dimensional inputs~\cite{Wu2026Channels}. With a qubit output and a qubit environment, Eq.~\eqref{eq:channel-capacity} covers arbitrary three- and four-dimensional inputs.

\emph{Conclusion and discussion.}---We proved strong superadditivity of EOF for any two-qubit marginal of a finite-dimensional bipartite state, with one qubit on each side. On tensor products, this yielded EOF additivity whenever one factor is a two-qubit state. We thus showed that Wootters' formula gives the exact entanglement cost of every two-qubit state. For a multiqubit state, Eq.~\eqref{eq:cost} gives the entanglement cost of each cross-system two-qubit marginal directly from Wootters' formula. Summing these costs over any matching gives a lower bound on the global preparation rate even when the pairs are correlated. The minimum global entanglement cost compatible with the selected pair marginals is the sum of the individual pair costs. We then determined the exact parallel LOCC simulation cost of every qubit-to-qubit channel from the EOF of its normalized Choi state. Finally, we derived a single-use formula for the energy-constrained classical capacity when the output and environment are qubits, including for arbitrary three- and four-dimensional inputs.

How far does strong superadditivity of EOF extend when one selected subsystem is a qubit and the dimension of the other increases? The qubit--qutrit case is a natural starting point for further investigation. Extending our theorem to qubit--qudit marginals would identify the entanglement cost of the corresponding states with their single-copy EOF. It would also yield single-use expressions for the parallel LOCC simulation cost of qubit-to-qudit and qudit-to-qubit channels and for the energy-constrained classical capacity of qubit-output channels with larger environments.
\par\endgroup

\begin{acknowledgments}
\emph{Acknowledgments.}---We set the research objectives and directed the proof strategy, using OpenAI Codex (GPT-6 Astra) to help develop and check the proof of Theorem~\ref{eof:strong-superadditivity} and to prepare the manuscript. We reviewed the proof of Theorem~\ref{eof:strong-superadditivity} step by step and take full responsibility for this work.
This work was supported by the National Science
Foundation of China under (Grants No. 12475009 and
No. 12075001), Anhui Provincial Key Research and
Development Plan (Grant No. 2022b13020004), Anhui
Province Science and Technology Innovation Project
(Grant No. 202423r06050004),  Anhui Provincial
University Scientific Research Major Project (Grant
No. 2024AH040008), Anhui Provincial Department of Industry and Information Technology (Grant No. JB24044), and the UK
Research and Innovation (Grant No. UKRI6536).
\end{acknowledgments}
\bibliography{references}

\input{Supplemental-Material}
\end{document}

%% file: Supplemental-Material.tex
\onecolumngrid
\clearpage
\makeatletter
\newcommand{\suppstartpage}{\zref@extractdefault{supp-start-page}{abspage}{5}}
\renewcommand{\thepage}{\ifnum\value{page}<\suppstartpage\relax\arabic{page}\else S\number\numexpr\value{page}-\suppstartpage+1\relax\fi}
\makeatother
\fontsize{10.95}{13.2}\selectfont
\phantomsection
\pdfbookmark[0]{Supplemental Material}{supplemental-material}
\makeatletter\zref@labelbyprops{supp-start-page}{abspage}\makeatother
\begin{center}
  \textbf{\large Supplemental Material for}\\[0.4em]
  \textbf{\large \manuscripttitle}\\[0.5em]
  Dong-Dong Dong, Xue-Ke Song, Liu Ye, Gerardo Adesso, and Dong Wang
\end{center}
\vspace{-1.5\baselineskip}
\setcounter{secnumdepth}{3}
\setcounter{section}{0}
\setcounter{equation}{0}
\renewcommand{\thesection}{S\arabic{section}}
\renewcommand{\thesubsection}{\thesection.\arabic{subsection}}
\renewcommand{\thesubsubsection}{\thesubsection.\arabic{subsubsection}}
\makeatletter
\renewcommand{\p@subsection}{}
\renewcommand{\p@subsubsection}{}
\makeatother
\renewcommand{\theequation}{S\arabic{equation}}
\renewcommand{\theHsection}{supp.\arabic{section}}
\renewcommand{\theHsubsection}{\theHsection.\arabic{subsection}}
\renewcommand{\theHsubsubsection}{\theHsubsection.\arabic{subsubsection}}
\renewcommand{\theHequation}{supp.\arabic{equation}}

\begingroup
{%
\section{Proof of Theorem~\ref{eof:strong-superadditivity} and derivation of Eq.~(\ref{eof:matching-bound})}
\label{app:eof-strong-superadditivity}
Lemma~\ref{app:match-entropy-lemma} shows that, for a pure global state,
suitable local measurements on the selected qubit pair reduce the
entanglement across the cut by at least the pair's EOF on average. We derive this
from a four-block entropy estimate and a two-qubit correlation
bound. Applying the convex roof and convexity on both marginals
proves strong superadditivity, from which additivity and the
two-qubit entanglement cost follow. Finally, iteration over a
matching and regularization give Eq.~\eqref{eof:matching-bound}.}

Throughout this section, $\rho_X$ denotes the reduction of the state
under consideration to $X$, and $S$ uses logarithms to base two.
For a two-qubit state $\omega$, $E_F(\omega)$ and
$\mathcal C(\omega)$ denote its EOF and concurrence across its two
qubits. Define
\begin{equation}
 W(c)=h_2\!\left(\frac{1+\sqrt{1-c^2}}2\right),\qquad
 h_2(q)=-q\log_2q-(1-q)\log_2(1-q),\qquad 0\le c,q\le1.
 \label{app:match-W}
\end{equation}
Wootters' formula gives $E_F(\omega)=W(\mathcal C(\omega))$
for a two-qubit state $\omega$~\cite{Wootters1998}.
We use $0\log_2 0=0$. The function $W$ is increasing on $[0,1]$,
and concurrence is normalized to one on a Bell state. Its homogeneous
extension to unnormalized operators is defined where needed below.

\subsection{Four-block entropy estimate}
Let $\rho=|\psi\rangle\langle\psi|$ be a pure state on
$A_1A'B_1B'$, where $A_1,B_1$ are qubits
and $A',B'$ have arbitrary finite dimensions $d,e$. In local bases of
the selected qubits its coefficient matrix across $A_1A'\mid B_1B'$ is
\begin{equation}
 V=\begin{pmatrix}M_{00}&M_{01}\\M_{10}&M_{11}\end{pmatrix},
 \qquad \rho_{A_1A'}=VV^\dagger,\qquad
 \operatorname{Tr}VV^\dagger=1.
\end{equation}
Each block $M_{xy}$ is $d\times e$, with $x,y\in\{0,1\}$
labeling the local outcomes with eigenvalues $(-1)^x$ and $(-1)^y$.
The branch $xy$ occurs with probability
$p_{xy}=\operatorname{Tr}M_{xy}M_{xy}^\dagger$. For $p_{xy}>0$,
the branch is pure on $A'B'$ and has marginal
$\rho_{A'}^{xy}=M_{xy}M_{xy}^\dagger/p_{xy}$ on $A'$.
The average branch entanglement is
$\sum_{xy}p_{xy}S(\rho_{A'}^{xy})$; a zero-probability term is defined to
be zero. The entropy loss is
\begin{equation}
 \Delta(V)=S(VV^\dagger)-\sum_{x,y=0}^1p_{xy}S(\rho_{A'}^{xy}).
 \label{app:match-loss}
\end{equation}
Write $\|X\|_p=(\operatorname{Tr}|X|^p)^{1/p}$ for the Schatten
norm, where $|X|=\sqrt{X^\dagger X}$. Split the matrix into its
even and odd blocks:
\begin{equation}
 \begin{gathered}
 U=\begin{pmatrix}M_{00}&0\\0&M_{11}\end{pmatrix},\qquad
 Q=\begin{pmatrix}0&M_{01}\\M_{10}&0\end{pmatrix},\\
 p_{\mathrm e}=\|U\|_2^2=p_{00}+p_{11},\quad
 p_{\mathrm o}=\|Q\|_2^2=p_{01}+p_{10}=1-p_{\mathrm e}.
 \end{gathered}
\end{equation}
The matrices $U+Q=V$ and $U-Q$ have the same singular values:
$U-Q=Z_dVZ_e$, where $Z_d=\operatorname{diag}(I_d,-I_d)$ and
$Z_e=\operatorname{diag}(I_e,-I_e)$.
Hanner's Schatten-norm inequality, proved for arbitrary complex
matrices and every $p\ge2$ in Ref.~\cite[Corollaries 1.1 and 1.5]{Heinavaara2025},
therefore yields
\begin{equation}
 2\|V\|_p^p\le
 (\|U\|_p+\|Q\|_p)^p+|\|U\|_p-\|Q\|_p|^p.
 \label{app:match-hanner}
\end{equation}
Rectangular matrices are covered by zero padding. No positivity
assumption is imposed on $V,U,Q$.

We spell out the derivative at $p=2$, since its direction and boundary
cases are essential. Temporarily write $s(\omega)=(\ln2)S(\omega)$
and $h_{\rm nat}=(\ln2)h_2$ for the natural-logarithm entropies, and set
$f(p)=\|V\|_p^p$, $u(p)=\|U\|_p$, $v(p)=\|Q\|_p$, and
$g(p)=((u(p)+v(p))^p+|u(p)-v(p)|^p)/2$.
For $p_{\mathrm e}p_{\mathrm o}>0$, Eq.~\eqref{app:match-hanner} gives
$f(p)\le g(p)$ for $p\ge2$, with $f(2)=g(2)=1$.
Hence $f'(2+)\le g'(2+)$; all derivatives below are taken from the right.
Direct differentiation gives
\begin{equation}
 \begin{aligned}
 f'(2)&=-\tfrac12s(VV^\dagger),\\
 u'(2)&=-\tfrac14\sqrt{p_{\mathrm e}}\,s(UU^\dagger/p_{\mathrm e}),\qquad
 v'(2)=-\tfrac14\sqrt{p_{\mathrm o}}\,s(QQ^\dagger/p_{\mathrm o}).
 \end{aligned}
\end{equation}
For example, writing
$\operatorname{Tr}|U|^p=p_{\mathrm e}^{p/2}
\operatorname{Tr}(UU^\dagger/p_{\mathrm e})^{p/2}$ makes the cancellation
of the normalization terms in $u'(2)$ explicit.
Put $q=1/2+\sqrt{p_{\mathrm e} p_{\mathrm o}}$. Differentiating the explicit powers
in $g$ contributes $(\ln2-h_{\rm nat}(q))/2$, whereas differentiating
$u,v$ contributes
$-[p_{\mathrm e} s(UU^\dagger/p_{\mathrm e})+p_{\mathrm o} s(QQ^\dagger/p_{\mathrm o})]/2$.
Since $h_2(q)=W(|p_{\mathrm e}-p_{\mathrm o}|)$, this proves
\begin{equation}
 S(VV^\dagger)\ge
 p_{\mathrm e} S(UU^\dagger/p_{\mathrm e})+p_{\mathrm o} S(QQ^\dagger/p_{\mathrm o})
 +W(|p_{\mathrm e}-p_{\mathrm o}|)-1.
 \label{app:match-parity}
\end{equation}
If $p_{\mathrm e}=p_{\mathrm o}$, then $u(p)-v(p)=O(p-2)$, so the term
$|u(p)-v(p)|^p$ has zero right derivative at two. If $p_{\mathrm e}=0$
or $p_{\mathrm o}=0$, one entire matrix vanishes and
Eq.~\eqref{app:match-parity} is a direct equality, with the
zero-weight term omitted. Fixed zero singular values contribute
identically zero to each norm power. Thus no full-rank assumption
is needed. Likewise, $p_{xy}S(\rho_{A'}^{xy})$ extends continuously by
zero, since it lies between zero and $p_{xy}\log_2\min(d,e)$.

The direct-sum entropy formula in Eq.~\eqref{app:match-parity} gives
\begin{equation}
 \Delta(V)\ge
 p_{\mathrm e} h_2(p_{00}/p_{\mathrm e})+p_{\mathrm o} h_2(p_{01}/p_{\mathrm o})
 +W(|p_{\mathrm e}-p_{\mathrm o}|)-1,
 \label{app:match-general-loss}
\end{equation}
again omitting zero-weight terms. Write $\omega=\rho_{A_1B_1}$ as
\begin{equation}
 \omega=\frac14\left[I\otimes I+\boldsymbol a\cdot\boldsymbol\sigma\otimes I
 +I\otimes\boldsymbol b\cdot\boldsymbol\sigma
 +\sum_{k,l=1}^3T_{kl}\sigma_k\otimes\sigma_l\right].
 \label{app:match-bloch}
\end{equation}
Here $\boldsymbol\sigma=(\sigma_x,\sigma_y,\sigma_z)$,
$\boldsymbol a,\boldsymbol b\in\mathbb R^3$ are the local Bloch
vectors, and $T$ is the real correlation matrix.
Choose unit measurement directions $\boldsymbol x\perp\boldsymbol a$
and $\boldsymbol y\perp\boldsymbol b$. Both single-qubit outcome
distributions are uniform. With $t=\boldsymbol x^{\mathsf T}T\boldsymbol y$,
\begin{equation}
 p_{00}=p_{11}=(1+t)/4,\qquad
 p_{01}=p_{10}=(1-t)/4.
\end{equation}
Consequently the first two terms of
Eq.~\eqref{app:match-general-loss} sum to one, including $|t|=1$,
and
\begin{equation}
 \Delta(V)\ge W(|\boldsymbol x^{\mathsf T}T\boldsymbol y|).
 \label{app:match-unbiased-loss}
\end{equation}
It remains to choose such directions with correlation at least
$\mathcal C(\omega)$. The following two-qubit lemma supplies this
step independently of the dimensions of $A',B'$.

\subsection{Transverse correlation and concurrence}
\begin{lemma}
\label{app:match-transverse}
In any fixed pair of local marginal eigenbases of a two-qubit state
$\omega$, including arbitrary choices for degenerate marginals,
\begin{equation}
 \mathcal C(\omega)\le
 2(|\omega_{00,11}|+|\omega_{01,10}|).
 \label{app:match-coherences}
\end{equation}
In particular, there exist unit vectors $\boldsymbol x\perp\boldsymbol a$
and $\boldsymbol y\perp\boldsymbol b$ such that
$|\boldsymbol x^{\mathsf T}T\boldsymbol y|\ge\mathcal C(\omega)$.
\end{lemma}
\begin{proof}
{Fix the marginal eigenbases and let $z=\omega_{00,11}$, $w=\omega_{01,10}$. For local transverse axes at azimuths $\phi,\theta$, the product of the Pauli observables flips both bits. Its expectation is $t=2\operatorname{Re}[z e^{i(\phi+\theta)}+w e^{i(\phi-\theta)}]$. The sum and difference of the angles can be chosen independently to cancel the phases of $z$ and $w$. Hence the maximum of $|t|$ is $2(|z|+|w|)$, including when either coherence vanishes. It suffices to prove Eq.~\eqref{app:match-coherences} in these fixed bases.}

\emph{Rank at most two.}
First suppose $\operatorname{rank}\omega\le2$. A pure marginal makes
$\omega$ a product state, so assume $\omega_{A_1}$ is invertible.
Define
\begin{equation}
 J=\tfrac12(\omega_{A_1}^{-1/2}\otimes I)\omega
                 (\omega_{A_1}^{-1/2}\otimes I),\quad
 F=\sqrt{2\omega_{A_1}},\quad
 |\chi\rangle=(F\otimes I)|\Phi^+\rangle.
\end{equation}
Here $|\Phi^+\rangle=(|00\rangle+|11\rangle)/\sqrt2$.
The state $J$ has trace one and first marginal $I/2$. It is the
normalized Choi state of a trace-preserving qubit channel $\Lambda$
of Choi rank at most two, and
$\omega=(\operatorname{id}\otimes\Lambda)(|\chi\rangle\langle\chi|)$.
Equivalently, $\omega=(F\otimes I)J(F^\dagger\otimes I)$.
This is an algebraic representation, not a probabilistic filter in
the entropy-loss protocol.

Up to input and output unitaries, the rank-two channel normal form
of Ref.~\cite[Theorem 12 and Eq.~(17)]{RuskaiSzarekWerner2002} is
\begin{equation}
 \boldsymbol r\longmapsto D\boldsymbol r+\kappa\boldsymbol e_3,
 \quad D=\operatorname{diag}(d_1,d_2,d_1d_2),\quad
 1\ge d_1\ge d_2\ge0,\quad
 \kappa^2=(1-d_1^2)(1-d_2^2).
 \label{app:match-channel}
\end{equation}
The signs of the first two entries can be made nonnegative by
Pauli rotations, and their axes can be interchanged. The input
unitary is absorbed into $|\chi\rangle$, and the output unitary
rotates the physical marginal eigenaxes along with the state.
Thus arbitrary prescribed axes for zero Bloch vectors remain covered.
Denote the Choi state in the canonical basis by $J_D$:
\begin{equation}
 J_D=\frac14\begin{pmatrix}
 1+\kappa+d_1d_2&0&0&d_1+d_2\\
 0&1-\kappa-d_1d_2&d_1-d_2&0\\
 0&d_1-d_2&1+\kappa-d_1d_2&0\\
 d_1+d_2&0&0&1-\kappa+d_1d_2
 \end{pmatrix}.
\end{equation}
The identities
$\sqrt{(1\mp d_1d_2)^2-\kappa^2}=d_1\mp d_2$
and Wootters' formula give $\mathcal C(J_D)=d_2$.
The original $J$ and $J_D$ are related by local unitaries, so
$\mathcal C(J)=\mathcal C(J_D)$.
For a positive operator $\eta$ with nonzero trace, define
$\mathcal C(\eta)=(\operatorname{Tr}\eta)
\mathcal C(\eta/\operatorname{Tr}\eta)$, with $\mathcal C(0)=0$.
For an invertible $F$, this homogeneous concurrence obeys
\begin{equation}
 \mathcal C((F\otimes I)\eta(F^\dagger\otimes I))
 =|\det F|\,\mathcal C(\eta).
 \label{app:match-filter}
\end{equation}
Indeed the pure-vector expression $2|\det X|$ for a coefficient
matrix $X$ has exactly this scaling. The invertible filter gives a
bijection between all pure decompositions on both sides, so the
convex-roof infima scale equally. This proves
Eq.~\eqref{app:match-filter} without a separate entanglement bound.

The reference Bloch vector of $|\chi\rangle$ is the same
$\boldsymbol a$ as that of $\omega$. Denote its input Bloch vector
by $\boldsymbol r$ and set
$c_0=\sqrt{1-\|\boldsymbol r\|^2}
=\mathcal C(|\chi\rangle\langle\chi|)$.
Its Schmidt decomposition supplies an orthogonal matrix $O$ such that
\begin{equation}
 \boldsymbol a=O\boldsymbol r,\qquad
 T_\chi=O[c_0I+(1-c_0)\widehat{\boldsymbol r}
                         \widehat{\boldsymbol r}^{\mathsf T}].
\end{equation}
Here $\widehat{\boldsymbol r}=\boldsymbol r/\|\boldsymbol r\|$
when $\boldsymbol r\ne0$; for $\boldsymbol r=0$, the expression
means $T_\chi=O$.
The matrix $O$ need not be a proper rotation. After the channel,
\begin{equation}
 \boldsymbol b=D\boldsymbol r+\kappa\boldsymbol e_3,\quad
 T=T_\chi D+\boldsymbol a(\kappa\boldsymbol e_3)^{\mathsf T},\quad
 \mathcal C(\omega)=c_0d_2.
 \label{app:match-channel-output}
\end{equation}
The last identity follows from Eq.~\eqref{app:match-filter}, since
$|\det F|=c_0$ and $\mathcal C(J)=d_2$.

If $c_0d_2=0$, the desired inequality is trivial. Otherwise $D$ is
invertible. If both Bloch vectors are nonzero, choose a unit
$\boldsymbol y\in\operatorname{span}(\boldsymbol e_1,\boldsymbol e_2)$
with $\boldsymbol r^{\mathsf T}D\boldsymbol y=0$ and put
$\boldsymbol x=OD\boldsymbol y/\|D\boldsymbol y\|$.
There is always such a $\boldsymbol y$, since this imposes one
homogeneous constraint in a two-dimensional plane. Then
$\boldsymbol x\perp\boldsymbol a$, $\boldsymbol y\perp\boldsymbol b$, and
\begin{equation}
 \boldsymbol x^{\mathsf T}T\boldsymbol y
 =c_0\|D\boldsymbol y\|\ge c_0d_2=\mathcal C(\omega).
 \label{app:match-ranktwo-correlation}
\end{equation}
\emph{Degenerate marginals in the rank-two case.}
Let $\boldsymbol n_A,\boldsymbol n_B$ be the
prescribed marginal eigenaxes when either marginal is degenerate.
If $\boldsymbol a=0$ and $\boldsymbol b\ne0$, then
$\boldsymbol r=0$, $c_0=1$, and $\boldsymbol b=\kappa\boldsymbol e_3$.
Choose $\boldsymbol y$ in the same plane subject to
$\boldsymbol n_A^{\mathsf T}OD\boldsymbol y=0$.
If $\boldsymbol a\ne0$ and $\boldsymbol b=0$, invertibility implies
$\boldsymbol r\parallel\boldsymbol e_3$; choose $\boldsymbol y$
in that plane perpendicular to $\boldsymbol n_B$.
In both cases the same formula for $\boldsymbol x$ proves
Eq.~\eqref{app:match-ranktwo-correlation} in the prescribed planes.
Finally, if both Bloch vectors vanish, then $\boldsymbol r=0$,
$\kappa=0$, $c_0=1$, and $d_1=1$. Choose a unit vector satisfying
$\boldsymbol y\perp\boldsymbol n_B$ and
$\boldsymbol n_A^{\mathsf T}OD\boldsymbol y=0$.
Two homogeneous constraints in $\mathbb R^3$ have a nonzero
solution. The least singular value of $D$ is now $d_2$, giving
Eq.~\eqref{app:match-ranktwo-correlation} once again.
This proves the lemma for rank at most two, with all degeneracies.

\emph{Reduction of general rank.}
Fix the two marginal eigenbases and choose local
diagonal phases so $z,w\ge0$ are real. Consider the compact convex
set of density operators with diagonal local marginals, real $z,w$,
and $z,w\ge0$. Trace one and the equalities impose seven real
linear constraints: four for the marginal off-diagonal entries and
two for the imaginary parts of $z,w$, in addition to trace.
When exactly one of $z,w$ is zero, preserving that zero adds one
real constraint. For a state of rank $r\ge3$ with at least one of
$z,w$ positive, the $r^2$-dimensional space of supported Hermitian
perturbations therefore contains a nonzero $H$ preserving all these
conditions and any vanishing member of $z,w$, since $r^2\ge9>8$.

Small perturbations in both directions preserve positivity. Let
$[s_-,s_+]$ be the maximal interval for which $\omega+sH$ remains
in the convex set. It is compact and $s_-<0<s_+$.
At either endpoint the rank drops or an additional positive one
of $z,w$ becomes zero; otherwise the interval could be extended.
The identity
\begin{equation}
 \omega=\frac{s_+}{s_+-s_-}(\omega+s_-H)
       +\frac{-s_-}{s_+-s_-}(\omega+s_+H)
 \label{app:match-rank-reduction}
\end{equation}
gives a convex decomposition preserving the common bases and signs.
Repeat on each unfinished branch. Rank never increases, vanishing
members of $z,w$ stay zero, and rank plus the number of positive entries
among $z,w$ strictly decreases. Hence the procedure terminates
after finitely many steps at rank-at-most-two states or states
with $z=w=0$.

\emph{The terminal case $z=w=0$.}
These states have the form
\begin{equation}
 \omega=\begin{pmatrix}
 q_{00}&\xi&\zeta&0\\
 \xi^*&q_{01}&0&-\zeta\\
 \zeta^*&0&q_{10}&-\xi\\
 0&-\zeta^*&-\xi^*&q_{11}
 \end{pmatrix},\qquad q_{xy}=\omega_{xy,xy}.
 \label{app:match-zero-corners}
\end{equation}
Local diagonal phases make $\xi,\zeta$ real simultaneously and
preserve the displayed form. Every $2\times2$ block is then
symmetric, so partial transposition on $B_1$ leaves the state
unchanged: $\omega^{T_{B_1}}=\omega\ge0$.
The two-qubit PPT criterion therefore implies that $\omega$ is
separable~\cite{Horodecki1996PPT}, and hence $\mathcal C(\omega)=0$.

\emph{Completion of the lemma.}
In the resulting finite decomposition $\omega=\sum_lq_l\omega_l$,
let $z_l=(\omega_l)_{00,11}$ and $w_l=(\omega_l)_{01,10}$.
All $z_l,w_l$ are real and nonnegative in the same fixed bases.
Concurrence convexity and the two terminal cases yield
\begin{equation}
 \mathcal C(\omega)\le\sum_lq_l\mathcal C(\omega_l)
 \le2\sum_lq_l(z_l+w_l)=2(z+w).
\end{equation}
Undoing the initial phases proves Eq.~\eqref{app:match-coherences}.
The local eigenvalues need not agree among the components; only
diagonality in the common bases is needed. This is why the
degenerate cases above were included explicitly.
\end{proof}

\subsection{Single-pair entropy loss}
{The preceding estimates give the quantitative loss
needed to separate a two-qubit pair from an arbitrary
finite-dimensional remainder.}
\begin{lemma}
\label{app:match-entropy-lemma}
Let $\rho=|\psi\rangle\langle\psi|$ be a pure state on
$A_1A'B_1B'$, where $A_1,B_1$ are qubits and $A',B'$ are arbitrary
finite-dimensional systems. There exist local rank-one projective
measurements on $A_1,B_1$ whose normalized conditional pure states
$|\psi_{xy}\rangle$ on $A'B'$, with $x,y\in\{0,1\}$,
occur with probabilities $p_{xy}$ and satisfy
\begin{equation}
 S(\rho_{A_1A'})\ge E_F(\rho_{A_1B_1})
       +\sum_{x,y}p_{xy}S(\rho_{A'}^{xy}).
 \label{app:match-entropy-loss}
\end{equation}
Here $\rho_{A'}^{xy}=\operatorname{Tr}_{B'}
|\psi_{xy}\rangle\langle\psi_{xy}|$, and zero-probability terms are omitted.
\end{lemma}
\begin{proof}
Choose the transverse measurement directions supplied by
Lemma~\ref{app:match-transverse} for $\rho_{A_1B_1}$.
Equation~\eqref{app:match-unbiased-loss} and monotonicity of $W$ give
$\Delta(V)\ge W(|\boldsymbol x^{\mathsf T}T\boldsymbol y|)
\ge W(\mathcal C(\rho_{A_1B_1}))=E_F(\rho_{A_1B_1})$,
which is the claimed inequality.
\end{proof}
The bases may depend on the selected two-qubit reduction. No single
measurement basis for all states or all branches is asserted.
For $|\phi\rangle_{A_1B_1}\otimes|\eta\rangle_{A'B'}$, every
conditional state on $A'B'$ is $|\eta\rangle$, so equality holds.
Thus the single-pair loss cannot be uniformly increased as a
function of that pair's concurrence alone.

\begingroup

\subsection{Strong superadditivity, additivity, and entanglement cost}
\label{app:eof-additivity}
\begin{proof}[Proof of Theorem~\ref{eof:strong-superadditivity} and its consequences]
For a pure state on $A_1A'B_1B'$, the conditional pure states
in Lemma~\ref{app:match-entropy-lemma} form an ensemble of
$\rho_{A'B'}$. Their average entanglement is therefore at least
$E_F(\rho_{A'B'})$. Thus
Eq.~\eqref{app:match-entropy-loss} proves strong superadditivity
for pure states.

For any decomposition $\rho=\sum_lq_l|\psi_l\rangle\langle\psi_l|$,
average the pure-state inequality and use convexity of EOF on each
marginal. Taking the infimum over decompositions yields
\begin{equation}
 E_F(\rho_{A_1A'\mid B_1B'})
 \ge E_F(\rho_{A_1B_1})+
 E_F(\rho_{A'B'}),
 \label{app:match-twoqubit-superadditivity}
\end{equation}
This proves Theorem~\ref{eof:strong-superadditivity} for arbitrary
finite-dimensional $A',B'$ and unrestricted correlations with the pair.

For any two-qubit state $\rho_{A_1B_1}$ and any finite-dimensional
bipartite state $\sigma_{A'B'}$, apply
Eq.~\eqref{app:match-twoqubit-superadditivity} to
$\rho\otimes\sigma$. The reverse inequality follows by taking
products of pure-state ensembles, so
$E_F(\rho\otimes\sigma)
=E_F(\rho)+E_F(\sigma)$.
Iteration gives $E_F(\rho^{\otimes k})=kE_F(\rho)$.
Using the regularized-EOF characterization of entanglement
cost~\cite{HaydenHorodeckiTerhal2001} therefore gives
$E_C(\rho)=E_F(\rho)$. No additivity assumption enters
the proof of strong superadditivity.
\end{proof}

\subsection{Matching bound}
\label{app:eof-matching}
\begin{proof}[Derivation of Eq.~\eqref{eof:matching-bound}]
Fix a matching $M$ across $A\mid B$ and let $\rho$ be any
multiqubit state, pure or mixed. Apply
Eq.~\eqref{app:match-twoqubit-superadditivity} to one selected
pair, then to the reduced state on the remaining qubits, and
continue through $M$. Disjointness ensures that tracing out one
selected pair leaves the other selected pair marginals unchanged.
Discarding the nonnegative EOF of the final remainder yields
\begin{equation}
 E_F(\rho_{A\mid B})
 \ge\sum_{(i,j)\in M}E_F(\rho_{A_iB_j}).
 \label{app:match-mixed}
\end{equation}
No independence assumption on the pairs is required.

Repeat this fixed physical-qubit matching in every one of $k$
tensor copies across $A^{\otimes k}\mid B^{\otimes k}$. The selected pair reductions are unchanged, so
\begin{equation}
 E_F(\rho^{\otimes k})
 \ge k\sum_{(i,j)\in M}E_F(\rho_{A_iB_j}).
 \label{app:matching-tensor-powers}
\end{equation}
Divide by $k$, take the regularization limit defining
$E_C$~\cite{HaydenHorodeckiTerhal2001}, and maximize over $M$.
The inequality $E_F(\rho_{A\mid B})\ge E_C(\rho_{A\mid B})$
follows from product-ensemble subadditivity of EOF.

\emph{Attainability.}
Let the pairs in a fixed matching have any prescribed density matrices
$\rho_{A_iB_j}$ and let the remaining systems have a separable state
$\tau$. Take the product of those factors. Repeated full additivity,
proved above, gives global EOF $\sum_{(i,j)\in M}E_F(\rho_{A_iB_j})$,
since $E_F(\tau)=0$. The same matching supplies this lower
bound on cost, and cost is bounded above by global EOF. Thus both
inequalities are equalities for arbitrary mixed pair factors as well.
One entangled pair suffices to show that the coefficient one is optimal.
\end{proof}
\endgroup
\endgroup

%% file: references.bib
@article{Wootters1998,
  author = {Wootters, William K.},
  title = {Entanglement of Formation of an Arbitrary State of Two Qubits},
  journal = {Phys. Rev. Lett.},
  volume = {80},
  pages = {2245--2248},
  year = {1998},
  doi = {10.1103/PhysRevLett.80.2245}
}

@article{ChristandlWinter2004,
  author = {Christandl, Matthias and Winter, Andreas},
  title = {``Squashed entanglement'': An additive entanglement measure},
  journal = {J. Math. Phys.},
  volume = {45},
  pages = {829--840},
  year = {2004},
  doi = {10.1063/1.1643788}
}

@article{Adesso2026GaussianEOF,
  author = {Adesso, Gerardo},
  title = {Optimality of {Gaussian} Entanglement of Formation},
  journal = {Phys. Rev. Lett.},
  year = {2026},
  note = {accepted 17 November 2026},
  doi = {10.1103/5jfn-ms77},
  url = {https://journals.aps.org/prl/accepted/10.1103/5jfn-ms77}
}

@article{Bennett1996,
  author = {Bennett, Charles H. and DiVincenzo, David P. and Smolin, John A. and Wootters, William K.},
  title = {Mixed-state entanglement and quantum error correction},
  journal = {Phys. Rev. A},
  volume = {54},
  pages = {3824--3851},
  year = {1996},
  doi = {10.1103/PhysRevA.54.3824},
}

@article{HaydenHorodeckiTerhal2001,
  author = {Hayden, Patrick M. and Horodecki, Micha{\l} and Terhal, Barbara M.},
  title = {The asymptotic entanglement cost of preparing a quantum state},
  journal = {J. Phys. A: Math. Gen.},
  volume = {34},
  pages = {6891--6898},
  year = {2001},
  doi = {10.1088/0305-4470/34/35/314},
}

@article{VidalDurCirac2002,
  author = {Vidal, Guifr\'e and D{\"u}r, Wolfgang and Cirac, J. Ignacio},
  title = {Entanglement Cost of Bipartite Mixed States},
  journal = {Phys. Rev. Lett.},
  volume = {89},
  pages = {027901},
  year = {2002},
  doi = {10.1103/PhysRevLett.89.027901}
}

@misc{Fang2026,
 author={Fang, Kun},
 title={Entanglement cost of quantum depolarization},
 year={2026}, howpublished={arXiv:2610.00187v1},
 url={https://arxiv.org/abs/2610.00187v1}
}

@article{Heinavaara2025,
  author = {Hein\"avaara, Otte},
  title = {Tracial joint spectral measures},
  journal = {Invent. Math.},
  volume = {239},
  pages = {505--526},
  year = {2025},
  doi = {10.1007/s00222-024-01309-6}
}

@article{RuskaiSzarekWerner2002,
  author = {Ruskai, Mary Beth and Szarek, Stanislaw and Werner, Elisabeth},
  title = {An analysis of completely-positive trace-preserving maps on {$M_2$}},
  journal = {Linear Algebra Appl.},
  volume = {347},
  pages = {159--187},
  year = {2002},
  eprint = {quant-ph/0101003},
  archivePrefix = {arXiv}
}

@article{Horodecki1996PPT,
  author = {Horodecki, Micha{\l} and Horodecki, Pawe{\l} and Horodecki, Ryszard},
  title = {Separability of mixed states: necessary and sufficient conditions},
  journal = {Phys. Lett. A},
  volume = {223},
  number = {1--2},
  pages = {1--8},
  year = {1996},
  doi = {10.1016/S0375-9601(96)00706-2}
}

@article{Shor2004,
 author={Shor, Peter W.}, title={Equivalence of additivity questions in quantum information theory},
 journal={Commun. Math. Phys.}, volume={246}, pages={453--472}, year={2004}, doi={10.1007/s00220-004-1071-1}
}

@article{Hastings2009,
 author={Hastings, Matthew B.}, title={Superadditivity of communication capacity using entangled inputs},
 journal={Nat. Phys.}, volume={5}, pages={255--257}, year={2009}, doi={10.1038/nphys1224}
}

@article{Pomeransky2003,
  author = {Pomeransky, A. A.},
  title = {Strong superadditivity of the entanglement of formation follows from its additivity},
  journal = {Phys. Rev. A},
  volume = {68},
  pages = {032317},
  year = {2003},
  doi = {10.1103/PhysRevA.68.032317}
}

@article{CornelioDeOliveira2010,
  author = {Cornelio, Marcio F. and de Oliveira, Marcos C.},
  title = {Strong superadditivity and monogamy of the {R\'enyi} measure of entanglement},
  journal = {Phys. Rev. A},
  volume = {81},
  pages = {032332},
  year = {2010},
  doi = {10.1103/PhysRevA.81.032332}
}

@misc{TangZhuBaiWang2026,
  author = {Tang, Ziao and Zhu, Chengkai and Bai, Ge and Wang, Xin},
  title = {Classical Capacity and Entanglement Cost of the Amplitude Damping Channel},
  howpublished = {arXiv:2609.28592},
  year = {2026},
  url = {https://arxiv.org/abs/2609.28592}
}

@misc{KrugerWerner2005,
  author = {Kr{\"u}ger, O. and Werner, R. F.},
  title = {Some Open Problems in Quantum Information Theory},
  howpublished = {arXiv:quant-ph/0504166},
  note = {Problem 7},
  year = {2005},
  url = {https://arxiv.org/abs/quant-ph/0504166}
}

@misc{PRLSupplementalMaterial,
  note = {See the Supplemental Material appended to this manuscript for the proof of Theorem~1 and the derivation of the matching bound. The Supplemental Material includes Refs.~\cite{RuskaiSzarekWerner2002,Horodecki1996PPT}.}
}

@misc{GaoAdesso2026Multimode,
  author = {Gao, Yuan and Adesso, Gerardo},
  title = {Entanglement cost and classical capacity of multimode {Gaussian} states and channels},
  year = {2026},
  eprint = {2611.22915},
  archivePrefix = {arXiv},
  primaryClass = {quant-ph}
}

@article{Berta2013Channels,
  author = {Berta, Mario and Brand{\~a}o, Fernando G. S. L. and Christandl, Matthias and Wehner, Stephanie},
  title = {Entanglement Cost of Quantum Channels},
  journal = {IEEE Trans. Inf. Theory}, volume = {59}, pages = {6779--6795},
  year = {2013}, doi = {10.1109/TIT.2013.2268533}
}

@article{MSW2004,
  author = {Matsumoto, Keiji and Shimono, Toshiyuki and Winter, Andreas},
  title = {Remarks on Additivity of the {Holevo} Channel Capacity and of the Entanglement of Formation},
  journal = {Commun. Math. Phys.}, volume = {246}, pages = {427--442},
  year = {2004}, doi = {10.1007/s00220-003-0919-0}
}

@article{Holevo1998Coding,
  author = {Holevo, A. S.},
  title = {Quantum coding theorems},
  journal = {Russian Math. Surveys},
  volume = {53},
  number = {6},
  pages = {1295--1331},
  year = {1998},
  doi = {10.1070/RM1998v053n06ABEH000091}
}

@misc{Fang2026GADC,
  author = {Fang, Kun}, title = {The classical capacity of generalized amplitude-damping channels and its strong-converse exponent},
  year = {2026}, eprint = {2609.14621v2}, archivePrefix = {arXiv}, primaryClass = {quant-ph}
}

@misc{Wu2026Channels,
  author = {Wu, Peixue}, title = {Classical communication over quantum channels: does entanglement help in nature?},
  year = {2026}, eprint = {2610.40585}, archivePrefix = {arXiv}, primaryClass = {quant-ph}
}

@article{KhatriSharmaWilde2020,
  author = {Khatri, Sumeet and Sharma, Kunal and Wilde, Mark M.},
  title = {Information-theoretic aspects of the generalized amplitude-damping channel},
  journal = {Phys. Rev. A},
  volume = {102},
  pages = {012401},
  year = {2020},
  doi = {10.1103/PhysRevA.102.012401}
}
